\RequirePackage{lineno}
\documentclass[pra,twocolumn,showpacs,floatfix,amsfonts,superscriptaddress]{revtex4-1}

\usepackage{bm}
\usepackage{graphicx}  
\usepackage{times}

\usepackage{amsmath}     
\usepackage{amsfonts}  
\usepackage{amssymb}  
\usepackage{dcolumn}
\usepackage{color}

\newcommand{\bra}[1]{\langle #1 | \,}
\newcommand{\ket}[1]{\, | #1 \rangle}

\newcommand{\be}{\begin{equation}}
\newcommand{\ee}{\end{equation}}
\newcommand{\bea}{\begin{eqnarray}}
\newcommand{\eea}{\end{eqnarray}}
\newcommand{\olap}[2]{\langle #1 | #2 \rangle}
\newcommand{\aver}[1]{\langle #1 \rangle}

\newcommand{\Int}{\mathbb{Z}}

\newcommand{\Hil}{\mathbb{H}}

\begin{document}
%\linenumbers
\title{Symmetries and security of a quantum-public-key encryption based on single-qubit rotations}
\author{U. Seyfarth}
\affiliation{Institut f\"ur Angewandte Physik, Technische Universit\"at Darmstadt, Darmstadt D-64289, Germany}
\author{G. M. Nikolopoulos}
\affiliation{Institute of Electronic Structure {\it \&} Laser, FORTH, P. O. Box 1385, Heraklion 71110, Crete, Greece}
\author{G. Alber}
\affiliation{Institut f\"ur Angewandte Physik, Technische Universit\"at Darmstadt, Darmstadt D-64289, Germany}
\date{\today}

\begin{abstract} 
Exploring the symmetries underlying a previously proposed encryption scheme which relies on 
single-qubit rotations, we derive an improved upper bound on the maximum information that an eavesdropper 
might extract from all the available copies of the public key. Subsequently, the robustness of 
the scheme is investigated in the context of attacks that address each public-key qubit 
independently. The attacks under consideration make use of projective measurements on single 
qubits and their efficiency is compared to attacks that address many qubits collectively and require 
complicated quantum operations.     
\end{abstract}

\pacs{03.67.Dd, 03.67.Hk}

\maketitle

\section{Introduction} 

Quantum-public-key cryptography, where the public keys are quantum-mechanical systems, is a largely unexplored area of problems. 
Various cryptographic primitives can be defined in this context (e.g., digital signatures, identification schemes, encryption schemes, etc) 
which aim at different goals (e.g., integrity, confidentiality, etc) \cite{qphGC01,BCGST02,ACJ06,Buh01,qphIM08,Got05,KKNY05,HKK08,Kak06,Nik08}. 
Of particular interest are quantum-public-key encryption (QPKE) schemes \cite{Got05,KKNY05,HKK08,Kak06,Nik08} which facilitate the 
communication between many users over insecure channels. Typically, a legitimate user participating in such a QPKE scheme has to choose a random 
secret (private) key, and prepare the public key in a state that is in accordance with the private key. 
Many copies of the public-key state can be created in this manner and become available to any potential sender in an authenticated manner, 
e.g. via a key-distribution center, whereas the corresponding private key is never revealed and is used by the receiver for decryption only. 
In a nutshell, QPKE combines the provable security of quantum-key distribution (QKD) protocols 
\cite{RevQKD} with the flexibility of conventional public-key 
encryption schemes, facilitating thus the {\em key distribution} and the {\em key management} in large networks \cite{Nik08,book1}. 
Key distribution and key management are crucial issues associated with the security and the efficient operation of large networks, 
and cannot be solved efficiently in the context of QKD (followed by a classical symmetric 
cryptosystem) or quantum direct communication (QDC) protocols such as \cite{PP1,PP2,PP3}.  
The main reason is that, by construction, the protocols of 
QKD and QDC are point-to-point protocols, and thus the total number of secure links and keys scales quadratically with the number of users in the network. 
This power law can be improved if the communications are performed via a key distribution center (KDC) which possesses all the secret keys. 
In this case, however, the center becomes an attractive target, while a compromised KDC renters immediately 
all communications insecure. In QPKE schemes on the other hand, 
the KDC deals with the public keys only, whereas the private keys are in posession of the legitimate users \cite{remark5}.
The study of QPKE schemes is also of fundamental importance for the field of quantum cryptography because of the quantum trapdoor 
one-way functions, which are essential ingredients not only  for the development of efficient encryption schemes, but also for many other cryptographic primitives 
(digital signatures, fingerprinting, zero-knowledge protocols, etc) \cite{Buh01, qphGC01, qphIM08, ACJ06, book1, book3}.   

The mere fact that in QPKE schemes many copies of the public keys become available, allows an eavesdropper to launch new 
strategies that go beyond QKD and QDC protocols (e.g., see \cite{NikIoa09}). Although the actual state of the public key is unknown to an adversary, the multiple copies, 
when processed judiciously, may reveal more information on this state than a single copy. Hence, a security analysis of a particular QPKE 
scheme has to address questions related to the lengths of the private and the public keys, as well as the number of public-key copies that 
can become available before the entire cryptosystem is compromised. Clearly, such questions are intimately connected to specific aspects of 
QPKE, which are not present neither in QKD nor in QDC protocols.    

The QPKE scheme of \cite{Nik08} is rather intuitive as it relies on 
single-qubit rotations. The public key consists of a number of qubits that are 
prepared at random and independently in some unknown state. 
A message can be encrypted in one of the public keys 
by rotating appropriately the corresponding qubit states and the resulting 
cipher-state is subsequently sent for decryption. 
Due to its simplicity, this scheme may serve as a theoretical framework for addressing 
questions pertaining to the power and limitations of QPKE as well as its robustness 
against various types of attacks. In this context it has been shown recently 
that any deterministic QPKE requires randomness in order to be secure
against a forward-search attack \cite{NikIoa09}. Furthermore, 
in contrast to the classical setting, a QPKE scheme can be used 
as a black box to build a new randomized bit-encryption scheme 
that is no longer susceptible to this attack.

Here we discuss for the first time 
a symmetry that underlies the scheme of \cite{Nik08} and that reduces 
considerably the information that an eavesdropper might extract 
from the copies of the public key. Subsequently, we analyze the 
security of the protocol against attacks that aim at the encrypted message 
and that rely on individual projective measurements on the qubits of the 
public key(s) and of the cipher state. It is shown that the performance 
of such attacks can be slightly worse than the performance of the 
forward-search attack \cite{NikIoa09} which requires complicated 
quantum transformations that are beyond today's technology. 

We like to emphasize, that discussions on the scheme of \cite{Nik08} with an appropriate
choice of the parameters also apply on a specific so-called ping-pong protocol \cite{PP2}
that pertains to the category of the so-called quantum direct communication (QDC) protocols.
The different context has to be taken into account to achieve meaningful statements.

%%%

This paper is organized as follows: In Sec.~\ref{sec2} basic aspects of the recently introduced
quantum-public-key protocol of \cite{Nik08} are summarized. The influence of symmetric
eavesdropping strategies on upper bounds of the probability for an eavesdropper to guess 
correctly the private key or the encrypted message are investigated in Sec.~\ref{Sec3}.
Security aspects of the private key are discussed in Sec.~\ref{sec3a} on the basis of Holevo's
bound. In Sec.~\ref{sec3b} an attack on encrypted messages is studied, which pertains to individual
projective measurements on the qubits involved. As a main result it is shown that Eve's success
probability converges to the value of one half exponentially with the numbers of qubits in which
the message is encrypted with a scale depending on the number of its publicly available copies
of the public key. Furthermore, it turns out that the success probability of this attack differs only slightly
from the already known optimal probability of successful state estimation by means of collective
measurements. In addition, as discussed in \ref{sec3c}, the resulting lower bound of the
security parameter of the public-key protocol is also close to the previously derived security
parameter of the forward-search attack of Ref. \cite{NikIoa09}.
Finally, in Sec. \ref{sec3d} a symmetry-test attack with projective measurements is explored,
which attacks the message directly and makes use of only a single copy of the public-key quantum
state and the corresponding cipherstate.

%%%%%%%%%%%%%%%%%
% Section 2
%%%%%%%%%%%%%%%%%

\section{The protocol}
\label{sec2}
For the sake of completeness, let us summarize briefly the main ingredients 
of the protocol proposed in \cite{Nik08}.  Each user participating in the cryptosystem 
generates a key consisting of a private part and a public part, 
as determined by the following steps. 
\begin{enumerate}
\item Choice of a random positive integer $n\gg 1$. Additional limitations on $n$ will be derived in the 
following section.

\item Choice of a random integer string ${\bf k}$ of length $N$ i.e.,  
${\bf k}=(k_1,k_2,\ldots,k_N)$. Each integer $k_j$ is chosen at random and independently from $\Int_{2^n}$, 
and thus it has a uniform distribution over $\Int_{2^n}$.
 
\item The classical key ${\bf k}$ is used for the preparation of the $N$-qubit public-key state 
\begin{subequations}
\label{public_key}
\be
\ket{\Psi_{\bf k}(\theta_n)}=\bigotimes_{j=1}^N\ket{\psi_{k_j}(\theta_n)}
\label{public_key_N}
\ee
where
\bea
\ket{\psi_{k_j} (\theta_n)} 
&\equiv& \cos \left( \frac{k_j \theta_{n}}{2} \right) 
\ket{0_{z}} + \sin \left(\frac{k_j\theta_{n}}{2} \right) 
\ket{1_{z}},\phantom{aa}
\label{public_key_j}
\eea
while $\{\ket{0_z},\ket{1_z}\}$ denote the eigenstates of the Pauli 
operator $\hat{\sigma}_z\equiv\ket{0_z}\bra{0_z}-\ket{1_z}\bra{1_z}$, which form an orthonormal basis in the Hilbert space of a qubit. 
The  Bloch vector associated with (\ref{public_key_j}) is given by 
${\bf R}_j(\theta_{n}) =\cos(k_j \theta_{n})\hat{z} + \sin(k_j \theta_{n})\hat{x}$ with $\hat{x}$, $\hat{z}$ denoting unit vectors 
and with
\be
\theta_n = \pi / 2^{n-1}
\ee
denoting the elementary angle of rotations around the axis with unit vector $\hat{y}$. 

\item The private (secret) part of the key is ${\bf k}$, while the public part is 
$\{n,N,\ket{\Psi_{\bf k}(\theta_n)}\}$. 
\end{subequations}
\end{enumerate}

Note that, since each $k_j$ is distributed uniformly and independently over $\Int_{2^n}$, the random 
state $\ket{\psi_{k_j}(\theta_{n})}$ is uniformly distributed over the 
set of states 
\be
\Hil^{(n)}=\{\ket{\psi_{k_j}(\theta_{n})}|k_j\in \{0,\ldots,2^n-1\}\}. 
\ee
The state of the $j$th public-key qubit $\ket{\psi_{k_j}(\theta_{n})}$ is 
known if the corresponding Bloch vector (or equivalently the 
angle $k_j\theta_{n}$) is known. The full characterization of the angle $k_j\theta_n$ requires $n$ bits of information. 

In general, a legitimate user should never reveal his private key, whereas he can produce at will as many copies of the public key as needed.  
The number of public-key copies $T^\prime$ \cite{remark7}, however,  should be kept sufficiently small relative to $n$ (the precise relation will be discussed 
in Sec. \ref{sec3a}), so that the map    
\be
\label{map}
{\bf k} \mapsto \{T^\prime~\textrm{copies of}~\ket{\Psi_{\bf k}(\theta_{n})}\}
\ee
is a quantum one-way function by virtue of Holevo's theorem \cite{Nik08,NikIoa09}. 
The one-way property of the map (\ref{map}) is essential for the definition of the
public-key encryption in the present framework. 

Suppose now that Bob wants to communicate a binary plaintext ${\bf m}$ to Alice. 
The users have agreed in advance on two encryption operators 
$\hat{\cal E}_{0}$ and $\hat{\cal E}_{1}$ for encryption of 
bit $''0''$ and $''1''$, respectively. The key point here is that the bits of the plaintext (message)   
are assumed to be encrypted independently on public qubits that have been prepared at 
random and independently (see discussion above). Hence, for the sake of simplicity and 
without loss of generality, we can focus on the encryption of a 
one-bit message  $m\in\{0,1\}$. As discussed in \cite{Nik08,NikIoa09}, in this case 
the protocol is not secure when the bit is encrypted on the state of a single 
qubit. However, it has been shown in the context of a forward-search attack, that 
the robustness of the protocol increases considerably  
if $m$ is encoded in a randomly chosen $s$-bit codeword ${\bf w}$ with 
Hamming weight of parity $m$ which is subsequently encrypted on $s$ public qubits \cite{remark6}. 
Correspondingly, the  analysis of the following section  pertains to a 
one-bit message, which is encrypted in the parity of an $s$-bit codeword with $s$ playing 
the role of a security parameter. 

For the encryption of the one-bit message $m\in\{0,1\}$, Bob chooses at random a codeword 
${\bf w}\equiv(w_1,w_2,\ldots,w_s)$ of parity $m$, and obtains an authenticated copy \cite{remark1} of Alice's public 
key ($T^\prime -1$ public keys still remain publicly available). 
The codeword is encrypted by applying independent successive 
encryption operations on the first $s$ public qubits. The resulting 
(quantum) ciphertext is thus the $s$-qubit state 
\be
\label{cipherstate}
\ket{X_{{\bf k}, m}(\theta_n)}=
\bigotimes_{j=1}^s\hat{\cal E}_{w_j}\ket{\psi_{k_j}(\theta_n)}=
\bigotimes_{j=1}^s\ket{\chi_{k_j,w_j}(\theta_n)},
\ee
to be referred to hereafter as cipherstate. In this spirit, for the encryption of an $L$-bit message 
requires a public-key of length $N\geq Ls$.  
The cipherstate is sent to Alice who can obtain the message by means of 
a decryption procedure whose details are not essential for our purposes 
in this work. We only note here the crucial property that the encryption operations do not depend 
on Alice's private key, but the decryption operators do.  Moreover, to allow  
for a simple decoding we assume that 
\be
\label{encryption}
\hat{\cal E}_{w_j}\ket{\psi_{k_j}(\theta_n)}\to \ket{\psi_{k_j}(\theta_n+w_j\pi)}, 
\ee 
for $w_j\in\{0,1\}$ \cite{remark8}.

The primary objective of an eavesdropper􏰁 (Eve) in the context of QPKE is to recover the plaintext 
from the cipher state intended for Alice. On the other hand, there is always a more ambitious objective pertaining to 
the recovery of the private key from Alice's public key. A cryptosystem is considered to be broken with accomplishment 
of either of the two objectives, but in the latter case the adversary has access to all of the messages sent to Alice (see also 
related discussion in \cite{Nik08,book1}). 
It is essential therefore to ensure secrecy of the private key, before we discuss the secrecy of a message.  
In  Sec.  \ref{sec3a}, we derive restrictions on the parameters $n$ and $T^\prime$ so that 
the map (\ref{map}) is a quantum one-way function, and thus the recovery of the private key from the public keys 
is prevented.  

As far as the encryption of the message (or equivalently the codeword) is concerned, 
 we note that, in view of Eqs. (\ref{public_key_j}) and (\ref{encryption}),  the two possible values of the $j$th bit of the 
 codeword $w_j\in\{0,1\}$ are essentially
encrypted in orthogonal eigenstates of a basis, which is rotated relative to the basis
$\{\ket{0_z},\ket{1_z}\}$ by an unknown angle $k_j \theta_{n}$. This means that the cipher-qubit state is
parallel ($w_j=0$) or antiparallel $(w_j=1)$ to the corresponding public-qubit state. 
Thus, in the following analysis we consider two different classes of eavesdropping strategies, 
which aim at the encrypted message. The first class involves attacks that explore the 
symmetry between the public-key state and the cipher state to reveal the message. 
The other class pertains to attacks that extract information on the public key 
(and thus on the basis on which the message has been encoded), so that 
the message can be recovered by means of a projective measurement on the 
estimated basis. Clearly, for this second class of attacks the probability 
of successful decryption is expected to increase with the information gain on the public-key 
state.

%%%%%%%%%%
% Section 3
%%%%%%%%%%

\section{Symmetric Eavesdropping Strategies}
\label{Sec3}
In a single run of the protocol the fixed quantities are the secret key ${\bf k}$ (and thus the public key), as well as  
the codeword ${\bf w}$. In general, for a given eavesdropping strategy, 
the probability of successful eavesdropping in a single run of the 
protocol $P(\textrm{suc}|{\bf k},{\bf w})$ differs from the corresponding
probability obtained by averaging over all possible values of ${\bf k}$, i.e., 
\bea
\bar{P}(\textrm{suc}|{\bf w})&=&\sum_{\bf k}P({\bf k})P(\textrm{suc}|{\bf k},{\bf w})\nonumber\\
&=&\frac{1}{2^{nN}}\sum_{\bf k}P(\textrm{suc}|{\bf k},{\bf w}),
\label{P_av_k}
\eea
where for the last equation we have used the fact that ${\bf k}$ is uniformly distributed over $\{0,1\}^{nN}$. 
The one-bit message $m$ is encoded at random on one of the $2^{s-1}$ possible $s$-bit codewords 
with parity $m$ (examples are given in \cite{Nik08,NikIoa09}). Hence, the conditional probability for the codeword 
${\bf w}$ to occur, given a particular value of $m\in\{0,1\}$, is  $P({\bf w}|m)=2^{-(s-1)}$. 
However, from the point of view of an adversary, both values of $m\in\{0,1\}$ are equally probable and thus 
$P({\bf w})= \sum_{m}P({\bf w}|m)2^{-1}=2^{-s}$ i.e.,  the codewords have a uniform distribution 
over $\{0,1\}^s$. Therefore, the eavesdropping strategies we are going to discuss 
are symmetric with respect to all possible codewords \cite{remark2}, and thus we also have 
$
\bar{P}(\textrm{suc})\equiv 2^{-s}\sum_{\bf w}P(\textrm{suc}|{\bf w})=P(\textrm{suc}|{\bf w}). 
$

\subsection{Eve's point of view}
\label{sec3a}
Our first task is to find out how much information Eve may extract from $\tau$ available copies 
of the $j$th public qubit, and investigate the conditions under which the security of the private key 
is guaranteed. 

From Eve's point of view, the state of the $j$th public qubit is uniformly distributed over $\Hil^{(n)}$, with the corresponding {\em a priori} probability  
being $2^{-n}$. Hence, the density operator describing the state of $\tau$ copies of the $j$th public qubit is  
\bea
\label{rho_prior_m}
\rho_{j,\rm prior}^{(\tau)}&=&
\frac{1}{2^n}
\sum_{k_j^\prime=0}^{2^n-1}\left [\ket{\psi_{k_j^\prime} (\theta_n)} \bra{\psi_{k_j^\prime} (\theta_n)}\right ]^{\otimes \tau}\nonumber\\
&=&\frac{1}{2^n}
\sum_{k_j^\prime=0}^{2^n-1}
\ket{\Phi_{k_j^\prime}^{(\tau)} (\theta_n)} 
\bra{\Phi_{k_j^\prime}^{(\tau)} (\theta_n)},
\eea
where  $\ket{\Phi_{k_j^\prime}^{(\tau)} (\theta_n)}:=\ket{\psi_{k_j^\prime} (\theta_n)}^{\otimes\tau}$.
In the space of 
$\tau$-qubit states we have $\tau+1$ different subspaces each of which is spanned by all 
${\cal B}(\tau,l)=\binom{\tau}{l}$ eigenstates with the same Hamming weight $l$, i.e. 
the same number of qubits which are in the state $\ket{1_z}$.  
Within one of these subspaces, say ${\cal S}_l$, we  can define the fully symmetric state 
\[
\ket{l}=\sum_{i=1}^{{\cal B}} \ket{i}_l/{\sqrt{\mathcal{B}(\tau,l)}},
\]
where the sum runs over all the $\tau$-qubit eigenstates with the same Hamming weight $l$. 
The problem can be formulated entirely in terms of these $(\tau+1)$-symmetric states 
$\{\ket{l}:l=0,1,\ldots, \tau\}$ \cite{remark3}.

Using Eq. (\ref{public_key_j}), we have  
\begin{subequations}
\label{Psi_tau}
\bea
\ket{\Phi_{k_j^\prime}^{(\tau)} (\theta_n)}
&=&\sum_{l=0}^{\tau} \sqrt{{\cal B}(\tau,l)}f_{\tau,l}(k_j \theta_n)\ket{l},
\eea
with 
\be
f_{\tau,l}(k_j \theta_n)=\left [\cos\left(\frac{k_j \theta_{n}}{2}\right)\right ]^{\tau-l}
\left[ \sin\left(\frac{k_j\theta_{n}}{2}\right)\right ]^{l}.
\ee
\end{subequations}
Thus the density operator of Eq. (\ref{rho_prior_m}) reads 
\begin{subequations}
\label{rho_tau}
\bea
\label{rho_pri_m_2}
\rho_{j,\rm prior}^{(\tau)}&=&
\sum_{l,l^\prime=0}^{\tau}
C_{l,l^\prime}
\ket{l}\bra{l^\prime}
\eea
with 
\bea
\label{C_llp}
C_{l,l^\prime}&=& \frac{1}{2^n}\sqrt{{\cal B}(\tau,l){\cal B}(\tau,l^\prime)} \sum_{k_j^\prime=0}^{2^n-1} 
f_{\tau,l}(k_j \theta_n)f_{\tau,l^\prime}^\star(k_j \theta_n).\phantom{aaw}
\eea
\end{subequations}
In the appendix \ref{app1} we provide additional information on the form of the {\em a priori}
density operator $\rho_{j,\rm prior}^{(\tau)}$ as well as on some observations regarding its rank
and eigenvalues. What we have so far, however, suffices to provide an upper bound on
the von Neumann entropy $S[\rho_{j,\rm prior}^{(\tau)}]$ for any values of $\tau$ and $n$. In particular, instead of
saying that $\tau$ copies of the $j$th public-key qubit are distributed, 
we can say that one copy of a larger $(\tau+1)$-dimensional system becomes 
publicly available. Hence, we have 
\be
\label{entropy_pri_1}
S[\rho_{j,\rm prior}^{(\tau)}]\leq\log_2(\tau+1). 
\ee

The state described in Eq. (\ref{rho_prior_m}) is a convex "classical" mixture of quantum 
states $\{\ket{\Phi^{(\tau)}_{k_j}(\theta_{n})}\}$ which are distributed with probabilities $p_j=2^{-n}$. Albeit pure, the states
$\ket{\Phi^{(\tau)}_{k_j}(\theta_{n})}$ are not mutually 
orthogonal. As a result the von Neumann entropy for the density operator $\rho_{j,\rm prior}^{(\tau)}$ is strictly smaller 
than the Shannon entropy of the corresponding probability distribution $H(p_j)=n$ \cite{book3}.
The Holevo bound restricts Eve's average information gain $I_{\rm av}$ on the unknown state for $\tau$ copies. 
In particular,  the information gain is upper bounded by $S[\rho_{j,\rm prior}^{(\tau)}]$, and
in view of inequality (\ref{entropy_pri_1}) we obtain the result
\be
I_{\rm av}\leq \log_2(\tau+1).
\ee
On the other hand, one still needs $n$ bits of information to characterize completely the state of the $j$th 
qubit (which of course implies knowledge on the private key as well). So, as long as  
\be
\label{holevo2} 
n\gg \log_2(\tau+1),
\ee
the one-way property of the map (\ref{map}) is guaranteed \cite{remark9}. Thus  
one can be confident that no matter what strategy Eve may choose, 
her information on each public-key qubit is very low. Despite the fact that 
Eve has almost no knowledge about the public key she may be able to decrypt 
an encrypted message successfully. This will be demonstrated in the next sections.

In closing, we would like to emphasize that in \cite{Nik08,NikIoa09} 
the symmetries underlying the particular encryption scheme have not been taken into account
and thus a larger upper bound on $I_{\rm av}$ was obtained suggesting that Eve can get up to 
$\tau$ bits of information from $\tau$ copies of the public key. However, this section demonstrates 
that the actual upper bound turns out to scale logarithmically with $\tau$
so that secrecy of the private key can be guaranteed already for significantly smaller values of $n$. 
Intuitively, this originates from the fact that the protocol restricts Eve by construction on the 
$(\tau+1)$-dimensional subspace of symmetric states for the $\tau$ copies of the $j$th public-key qubit.
In appendix \ref{app1} we provide a tighter upper bound on Eve's information gain based on basic properties
of the eigenvalues of $\rho_{j,\rm prior}^{(\tau)}$.

\begin{figure*}
\includegraphics[scale=0.7]{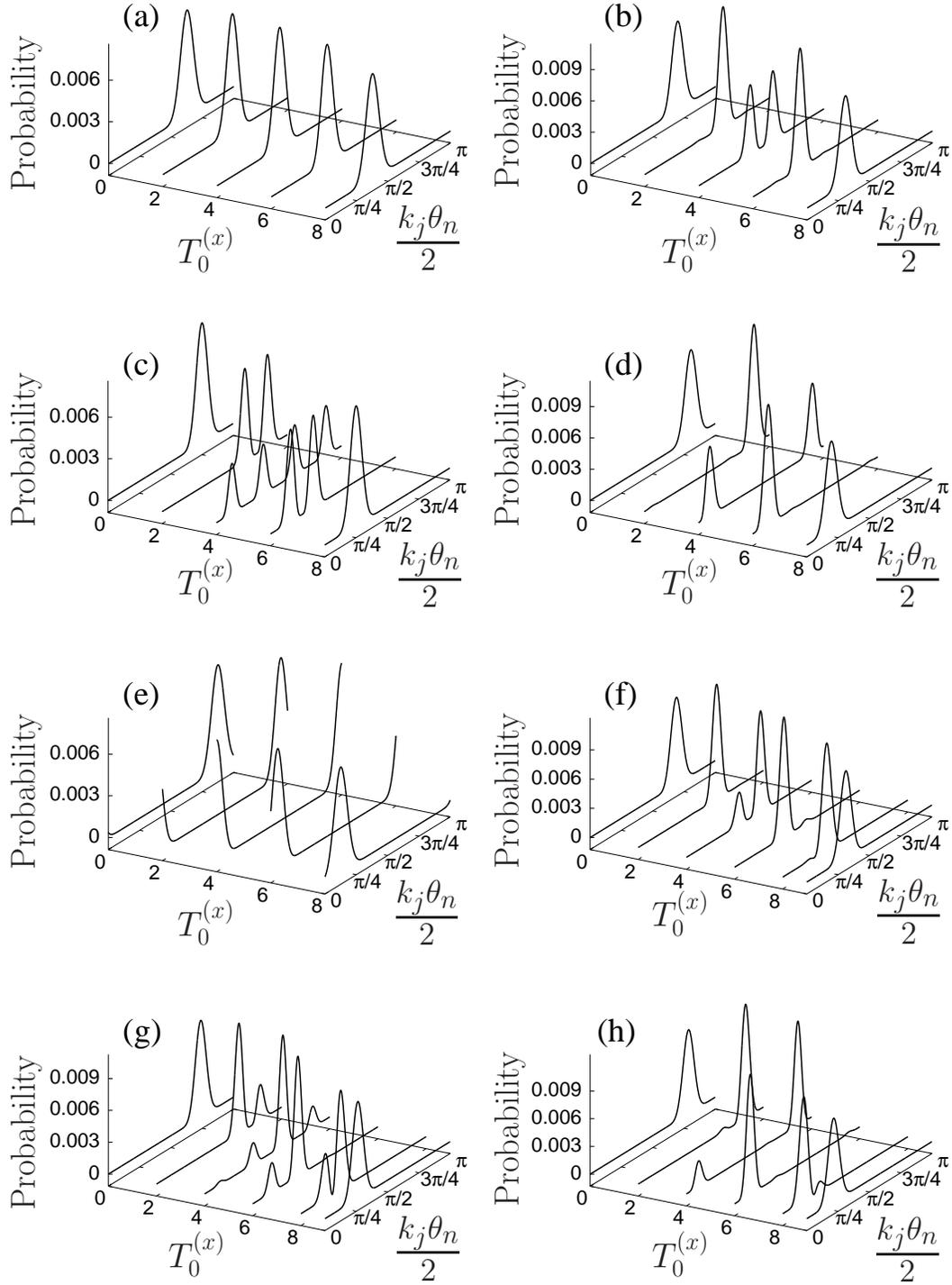}
\caption{
A posteriori probability distributions (given by Eqs. \ref{eq:apost}) for $T=8$ (a-e), $T=9$ (f-h), and  various events $\{T_0^{(z)},T_0^{(x)}\}$:
(a) $T_0^{(z)}=0$; (b,f) $T_0^{(z)}=2$; (c,g) $T_0^{(z)}=4$; (d,h) $T_0^{(z)}=6$; (e) $T_0^{(z)}=8$.
}
\label{fig1}
\end{figure*}

\subsection{Incoherent Projective Measurements}
\label{sec3b}
Eve knows that all of the qubit states lie on the $x-z$ plane of the Bloch sphere. 
Thus, she may try to deduce the message by means of projective measurements on 
the cipherstate as well as on all of the remaining 
$(T^\prime-1)$ copies of the public key \cite{remark4}. In the following, we assume that 
each qubit of the public key or of the cipher is measured independently. 
Indeed, given that the random state of each public-key qubit is chosen independently 
and that it is distributed uniformly over $\Hil^{(n)}$, it is reasonable to assume that 
there are no hidden patterns that Eve can take advantage of by attacking many qubits collectively.  
  
One possible strategy for Eve is to obtain an estimate of the public-key state (\ref{public_key})
by measuring half of the public keys on the (eigen)basis $\{\ket{0_z},\ket{1_z}\}$ of
the Pauli operator $\hat{\sigma}_{z}$ and the other half on the (eigen)basis $\{\ket{0_x},\ket{1_x}\}$ of the Pauli 
operator $\hat {\sigma}_x\equiv\ket{0_z}\bra{1_z}+\ket{1_z}\bra{0_z}$. 
In this way she can obtain an estimation on the $j$th public-qubit state
or equivalently on its Bloch vector ${\bf R}_j$.    
It should be emphasized that such an attack essentially aims at the private 
key which, by construction, is in one-to-one correspondence with the public key. 
Although, condition (\ref{holevo2}) restricts Eve's information gain on the private key to 
negligible values, it cannot guarantee secrecy of the encrypted message. 
Hence, in an attempt to reveal the message she can measure the cipherstate on a basis defined by 
her guess on the corresponding public-qubit state. The main purpose of this section is to analyze this attack. 

Since all public-key qubits are equivalent and independent, let us start by focusing on 
one of them, i.e., the $j$th qubit which is measured in the basis $b\in\{z,x\}$ with $b=z(x)$ 
referring to the eigenbasis of the operator $\hat{\sigma}_z(\hat {\sigma}_x)$. 
The two possible outcomes of these measurements are "0" and "1" and they occur with probabilities 
\bea
p_{j,0}^{(b)}(k_j)=\cos^2\left ( \beta\frac{\pi}{4}-\frac{k_j\theta_{n}}2\right ),\quad p_{j,1}^{(b)}=1-p_{j,0}^{(b)}.
\eea
In this equation, $\beta\in\{0,1\}$ with 
the correspondences $b=z\rightarrow \beta=0$ and $b=x\rightarrow \beta=1$.
Without loss of generality let us also assume that $T^\prime-1=2T$ \cite{remark4}, so 
that $T$ measurements are performed on the basis $b$. 
Let $T_0^{(b)}$ denote the number of outcomes "0" from measurements 
in the $b$ basis. In a single run of the protocol Eve obtains a particular set of outcomes $\{T_0^{(z)},T_0^{(x)}\}$
out of $T^2$ different possible combinations. 
We will first discuss how much information she can obtain about the public-qubit state (or equivalently the private key).

\subsubsection{Information gain on the public-qubit state}
\label{sec3b1}
The {\em a posteriori} probability for the $j$-th qubit state is given by Bayes law 
\begin{subequations}
\label{eq:apost}
\bea
p_j(k_j^\prime|T_0^{(z)},T_0^{(x)})=\frac{q_j(T_0^{(z)},T_0^{(x)}|k_j^\prime)}{2^nq(T_0^{(z)},T_0^{(x)})}.
\eea
The probability for the outcome $\{T_0^{(z)},T_0^{(x)}\}$ to occur given the 
input state $\ket{\psi_{k_j^\prime}(\theta_n)}$ is
\bea
q_j(T_0^{(z)},T_0^{(x)}|k_j^\prime)&=&
\binom{T}{T_0^{(z)}}\binom{T}{T_0^{(x)}}
\times\nonumber\\ & &\times\prod_b
\left [p_{j,0}^{(b)}(k_j^\prime) \right ]^{T_0^{(b)}}  
\left [p_{j,1}^{(b)}(k_j^\prime)\right ]^{T-T_0^{(b)}},\nonumber\\
\eea
and 
\bea
q(T_0^{(z)},T_0^{(x)})=\frac{1}{2^n}
\sum_{k_j^\prime=0}^{2^{n}-1}q_j(T_0^{(z)},T_0^{(x)}|k_j^\prime).
\eea
\end{subequations}
A sample of {\em a posteriori} probability distributions is depicted in Fig. \ref{fig1}, for $T=8$, $T=9$, and various events $\{T_0^{(z)},T_0^{(x)}\}$. 
Different public-qubit states may give rise to a certain combination $\{T_0^{(z)},T_0^{(x)}\}$ albeit with different 
probabilities. Hence, given a particular combination of "0" outcomes in the two bases, the conditional 
{\em a posteriori} probability distribution exhibits peaks for public-qubit states (as determined by $k_j \theta_{n}$),
which are consistent with the particular event under consideration. 
 
\begin{figure}
\includegraphics[scale=0.33]{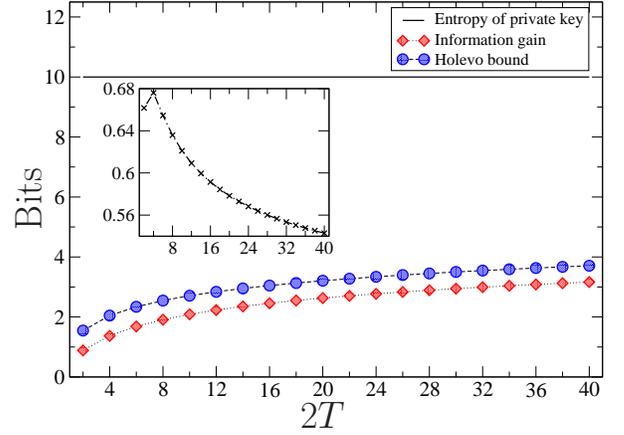}
\caption{(Color online) Entropy of {\em a priori} probability distribution (=entropy of private key), Holevo bound and information gain as 
functions of the number of public-key copies $2T$ that become available. The value of $n$ affects considerably 
the {\em a priori} probability distribution. The inset shows the difference between the Holevo bound and the information gain.}
\label{fig2}
\end{figure}

Eve's information gain is given by the difference of the Shannon entropies of the distributions before and after the measurements, 
i.e.,  
% \begin{widetext}
\bea
I_{\rm av}&=&H_{\textrm{prior}}-
\aver{H_{\textrm{post}}}\nonumber\\
&=&n+\sum_{T_0^{(z)}}\sum_{T_0^{(x)}}q(T_0^{(z)},T_0^{(x)})\times\nonumber\\ & &\times
\sum_{k_j^\prime=0}^{2^n-1}p_j(k_j^\prime|T_0^{(z)},T_0^{(x)})\log[p_j(k_j^\prime|T_0^{(z)},T_0^{(x)})]\phantom{aaa}
\eea
% \end{widetext}
where we have summed over all possible outcomes for a given state. The entropy of 
the {\em a priori} uniform probability distribution is equal to the entropy of the private-key bit $k_j$. 
As depicted in Fig. \ref{fig2}, this information gain is slightly below the Holevo bound 
of Eq. (\ref{entropy_pri}) for $\tau=2T$, which is tighter than the bound of Eq. (\ref{entropy_pri_1}).
It is worth mentioning that although the information gain depends 
weakly on $n$ the Holevo bound does not. In the subsequent discussion the choices of 
$n$ and $T$ are such that the inequality (\ref{entropy_pri}) and thus also inequality (\ref{entropy_pri_1})
are satisfied for $\tau=2T$. 

\subsubsection{Probability of correct guessing the message}

As we have seen in the previous subsection, a particular outcome $\{T_0^{(z)},T_0^{(x)}\}$ of a single run of the protocol allows Eve to update her knowledge on the public-qubit state she may have been given. 
From her point of view the {\em a posteriori} state pertaining to $\tau$ public-key copies is given by
\bea
\label{rho_post_m}
\rho_{j,\rm post}^{(\tau)}(T_0^{(z)},T_0^{(x)})&=&\sum_{k_j^\prime=0}^{2^n-1}p(k_j^\prime|T_0^{(z)},T_0^{(x)})\times\nonumber\\& &\times
\ket{\Phi_{k_j^\prime}^{(\tau)} (\theta_n)} 
\bra{\Phi_{k_j^\prime}^{(\tau)} (\theta_n)}.
\eea

Tracing out $\tau-1$ copies, we obtain for the single-copy density operator the expression
\bea
\rho_{j,\rm post}^{(1)}=\sum_{k_j^\prime}
p(k_j^\prime|T_0^{(z)},T_0^{(x)})\ket{\psi_{k_j^\prime}(\theta_n)}\bra{\psi_{k_j^\prime}(\theta_n)}\label{eqn:rhopost1}
\eea
and the corresponding (estimated) Bloch vector
\bea
\tilde{\bf R}_j&=&\sum_{k_j^\prime} p(k_j^\prime|T_0^{(z)},T_0^{(x)}) 
[\cos(k_j^\prime\theta_n) \hat{z} +  \sin(k_j^\prime \theta_{n}) \hat{x}]\phantom{aaa}
\eea
with $||\tilde{\bf R}_j||\neq 1$.

Recall now that the one-bit message $m$ is encoded in the parity of an $s$-bit codeword 
${\bf w}$ which is subsequently encrypted on $s$ public qubits. 
Let us calculate first Eve's probability to recover the bit $w_j$ in a single run of the protocol
by measuring the corresponding cipher qubit in the basis defined by $\tilde{\bf R}_j$. For the
particular encryption under consideration (see Sec. II) her probability of success is 
$P(\textrm{suc}|w_j, k_j, T_0^{(z)}, T_0^{(x)})=\cos^2(\Omega_j/2)$ with $\Omega_j$ denoting the angle between the actual Bloch vector 
${\bf R}_j$ and its estimation $\tilde{\bf R}_j$. Hence, we obtain
\bea
P(\textrm{suc}|w_j, k_j, T_0^{(z)}, T_0^{(x)})=\frac{1}2+\frac{\tilde{\bf R}_j\cdot{{\bf R}_j}}{2||\tilde{\bf R}_j||}\quad\nonumber\\
=\frac{1}2+\frac{1}{2||\tilde{\bf R}_j||}\sum_{k_j^\prime} p(k_j^\prime|T_0^{(z)},T_0^{(x)}) \cos[(k_j^\prime-k_j)\theta_{n}]
\eea
with $ {{\bf R}_j}$ defined in Sec. \ref{sec2}.
For a given public-qubit state various outcomes may occur albeit with different probabilities
\bea
P(\textrm{suc}|w_j, k_j)=\sum_{T_0^{(z)}} \sum_{ T_0^{(x)}}& P(\textrm{suc}|w_j,k_j, T_0^{(z)}, T_0^{(x)})\times\nonumber\\&\times q( T_0^{(z)}, T_0^{(x)}|k_j). %\\
\eea

\begin{figure}
\includegraphics[scale=0.33]{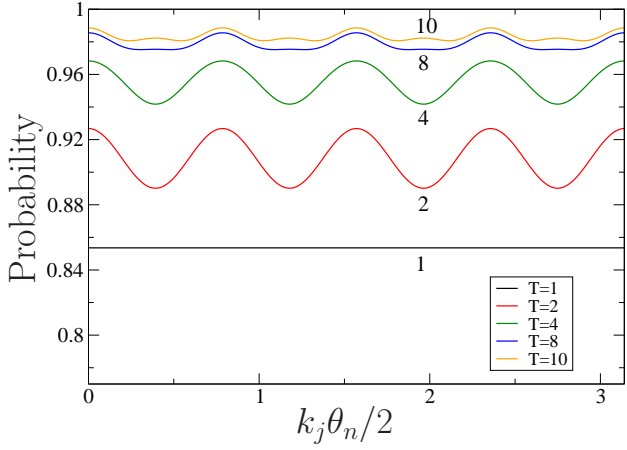}
\caption{(Color online) Conditional probability $P(\textrm{suc}|w_j, k_j)$ for $n=10$ and various values of $T$.}
\label{fig3}
\end{figure}

The typical behavior of $P(\textrm{suc}|w_j, k_j)$ with $k_j$ (or equivalently $k_j \theta_{n}$)  
is depicted in Fig. \ref{fig3} where we have an oscillation around the mean value
\bea
\bar{P}(\textrm{suc}|w_j)=\frac{1}{2^n}\sum_{k_j}P(\textrm{suc}|w_j, k_j). 
\eea 
As we increase the number of public-key copies the amplitude of the oscillations becomes smaller and the mean value increases. 
In particular, we find that for $T>1$
\bea
\bar{P}(\textrm{suc}|w_j)\lesssim 1-\frac{1}{6T}:=U(T).
\label{U_of_T}
\eea 

\begin{figure}
\includegraphics[scale=0.33]{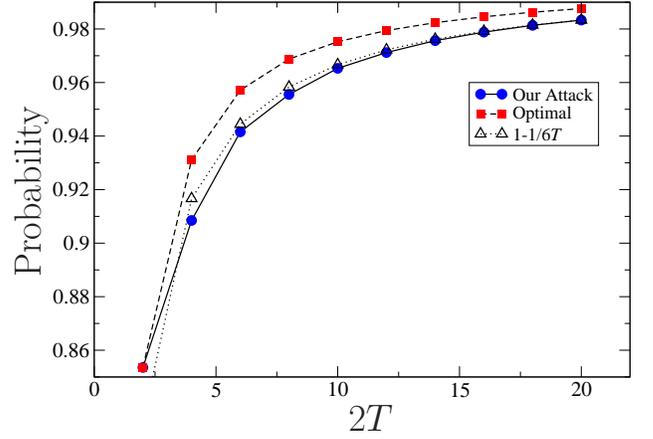}
\caption{(Color online) Conditional probability $P(\textrm{suc}|w_j)$ for $n=10$ and various values of $T$.}
\label{fig4}
\end{figure}

As depicted in Fig. \ref{fig4}, this performance is very close to the {\em optimal} probability of successful state estimation by means 
of collective measurements \cite{Der98} 
\bea
\bar{P}_{\rm opt}(\textrm{suc}|w_j)=\frac{1}2+\frac{1}{2^{2T+1}}\sum_{i=0}^{2T-1} \sqrt{\binom{2T}{i}\binom{2T}{i+1}}
\eea 
which scales like
\bea
\bar{P}_{\rm opt}(\textrm{suc}|w_j)\sim 1-\frac{1}{8T}.
\eea 
Bagan {\em et al.} \cite{Bag02} have demonstrated that this upper bound can be saturated by means of individual measurements
and our attack has similarities to their approach. Finally, for our subsequent discussion it is worth keeping in mind that
$P(\textrm{suc}|w_j)$ does not depend on the actual value of the bit $w_j$ i.e., $P( \mathrm{suc} | w_j=0)=P( \mathrm{suc} | w_j=1)$.

Up to now our results are referring to one bit of the codeword only and our task is to 
obtain the probability of success in guessing correctly the bit-message $m$ from the $s$-bit 
codeword ${\bf w}$. Since the message is encoded on the parity of the codeword, 
Eve succeeds even if she fails to predict correctly 
$\alpha$ out of $s$ bits with $\alpha$ even. Instead of considering her probability of success in a single run 
of the protocol, which is a rather complicated task, we concentrate in the following on her probability 
of success averaged over all possible public-qubit states (or equivalently private keys ${\bf k}$). 
As depicted in Fig. \ref{fig3},  for large $T$ the amplitude of the oscillations is at least an order of magnitude 
smaller than the mean. Hence, any conclusions based on the average probability of success are also expected to 
apply with good accuracy to a single run of the protocol. 
Since each bit of the codeword is encrypted separately in independently prepared public qubits, the averaging 
over all possible values ${\bf k}$ is straightforward. Thus, one obtains for the average 
probability of successful eavesdropping for a given message $m$ and codeword ${\bf w}$
\begin{subequations}
\label{P_s_av1}
\bea
\bar{P}_s( \mathrm{suc}|{m},{\bf w}) = \sum_{\substack{\alpha=0\\ {\rm even}}}^s \binom{s}{\alpha}& [1-\bar{P}( \mathrm{suc}| w_j)]^{\alpha} \times \nonumber\\&\times
[\bar{P}( \mathrm{suc} | w_j)]^{s-\alpha}.\label{eqn:Ps}
\eea
Averaging over all possible equally probable codewords and messages we finally find
\bea
\bar{P}_s( \mathrm{suc})=\bar{P}_s( \mathrm{suc}|{m},{\bf w}).
\eea 
\end{subequations}

\begin{figure}
\includegraphics[scale=0.33]{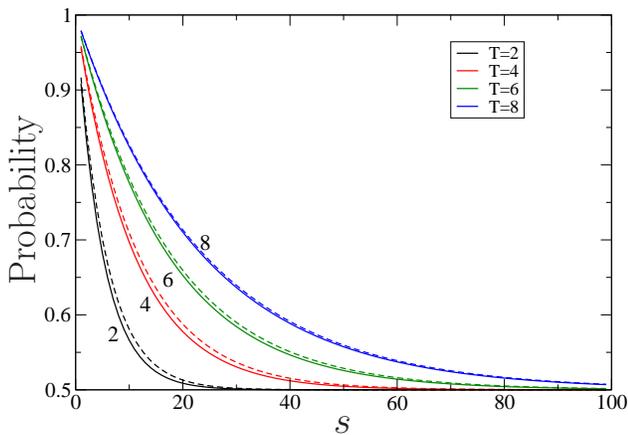}
\caption{(Color online) Average probability of success $\bar{P}_s( \mathrm{success})$ as a function of codeword length $s$, for $n=10$ and various values of $T$. 
The solid lines are numerical results obtained from Eqs. (\ref{P_s_av1}), whereas the dashed lines are for the upper bound 
defined in Eq. (\ref{Ubound}).}
\label{fig5}
\end{figure}

In Fig. \ref{fig5}, $\bar{P}_s( \mathrm{suc})$ is depicted as a function of the codeword length $s$ for various numbers of public-key copies (solid lines). 
Clearly, the average probability of success decreases with increasing $s$ whereas this drop becomes slower and 
slower as we increase the number of public-key copies.  For $T>1$ a rather tight upper bound for $\bar{P}_s( \mathrm{suc})$ 
is given by the expression
\bea
\frac{1}2+\frac{1}2\left ( 1-\frac{1}{3T}\right )^s
\label{Ubound}
\eea
which is also plotted in Fig. \ref{fig5} with dashed lines. A sketch of the proof of this upper bound is provided in Appendix \ref{app2}.

Now, let us assume that the users participating in the protocol have agreed in advance on a security parameter $\varepsilon\ll 1$ so that Eve's probability 
of success $\bar{P}_s( \mathrm{suc})$ has to fulfill the relation $\bar{P}_s( \mathrm{suc})\leq 1/2+\varepsilon$. This implies that the message bit $m$ 
has to be encrypted in 
\bea
s\geq \left |\frac{1+\log_2(\varepsilon)}{\log_2\left ( \frac{3T-1}{3T}\right )}\right |
\label{Lbound1}
\eea
qubits which is always fulfilled if
\bea
s\geq 3T |1+\log_2(\varepsilon)|. 
\label{Lbound2}
\eea

\subsection{Comparison to the forward-search attack}
\label{sec3c}
The robustness of the present public-key encryption scheme against a forward-search attack based on a symmetry test 
in which Eve compares the cipher state with the public-key state is discussed in Ref. \cite{Nik08,NikIoa09}. 
The symmetry test of Ref. \cite{Nik08,NikIoa09} takes into account all the copies of the public keys but  
in contrast to the attacks discussed here it requires rather 
complicated quantum operations and gates, such as Fourier transformations and permutations on large numbers of qubits.  
Due to the nature of the attack the probability for successful eavesdropping does not vary from run to run and 
the probability for an eavesdropper to deduce the parity of the $s$-bit codeword and hence the message from the cipherstate 
is given by \cite{NikIoa09}
\bea
\bar{P}_s(\textrm{suc})=\frac{1}2+\frac{1}2\left ( 1-\frac{1}{2T}\right )^s.
\label{swap}
\eea
It is rather surprising how close this exact expression is to the upper bound (\ref{Ubound}), which is slightly below the 
optimal probability of success. For a given 
security threshold $\varepsilon$ the length of the codeword has to satisfy 
\bea
s\geq T|1+\log_2(\varepsilon)|. 
\label{Lbound3}
\eea
which differs from Eq. (\ref{Lbound2}) by a factor of three only. 

\subsection{A symmetry-test attack with projective measurements}
\label{sec3d}
In contrast to the previous attack we will consider here an attack which aims 
directly at the message rather than the private key and makes use of  
one copy of the public-key state and the cipherstate only. Eve pairs up 
the corresponding qubits of the public key and the cipher state i.e., 
the $j$th pair pertains to the $j$th qubits. The qubits of the $j$th pair
are projected independently onto the same randomly chosen 
eigenbasis $\{\ket{0_{\varphi_j}}\},\ket{1_{\varphi_j}}\}$ where 
\bea
\ket{\zeta_{\varphi_j}}&=&(-1)^{\zeta}\cos\left (\frac{\varphi_j}{2}
\right )\ket{0_z}+\sin\left (\frac{\varphi_j}{2}\right )\ket{1_z}
\eea
and $\varphi_j$ is uniformly distributed over $[0,2\pi)$. 
The probability of correct guessing either of the qubits is given by 
\bea
F(k_j \theta_{n},\varphi_j)\equiv|\olap{\psi_{k_j}(\theta_n)}{\zeta_{\varphi_j}}|^2=\cos^2\left (
\frac{k_j \theta_{n}-\varphi_j}2
\right).\phantom{a}
\eea
However, since for a fixed value of $k_j$ the angle $\varphi_j$ is chosen at random, we can introduce a 
new random variable $\omega_{j,n}\equiv k_j \theta_{n}-\varphi_j$ uniformly distributed over the interval $[0,2\pi)$.
For later convenience let us also denote the number 
of wrong outcomes for the $j$th pair by $e_j$ with $0\leq e_j\leq 2$. 
As discussed in the last paragraph of Sec. \ref{sec2}, the question 
that Eve has to answer is whether the states of the qubits 
in the $j$th pair are parallel or antiparallel. She obtains the 
correct answer if the outcomes of the measurements on the corresponding 
two qubits are either both correct $(e_j=0)$ or both wrong $(e_j=2)$. 
Thus, the probability of success in a single run of this protocol is given by
\be
P(\textrm{suc}|w_j,k_j)=[F(\omega_{j,n})]^2+[1-F(\omega_{j,n})]^2.
\label{psU_1}
\ee
If the one-bit message is encoded in the parity of an $s$-bit codeword which is subsequently encrypted on $s$ qubits, 
Eve's strategy succeeds provided the total number of incorrect outcomes $e=\sum_{j=1}^s e_j$ is an even integer 
(e.g., see Table \ref{tab:1} for $s=2$). 
The total probability of success in a single run can be obtained by means of an iteration of the form (\ref{Qi}), where 
$Q^{(s)}$ is a multivariable function, i.e., $Q^{(s)}(\omega_{1,n},\ldots \omega_{s,n})\equiv P_s(\textrm{suc}|{\bf k}, {\bf w})$).
Hence, Eve's probability of success in getting the correct parity and thus the correct message
consists of two parts pertaining to possible combinations of outcomes from a single pair and the remaining $s-1$ pairs. 
More precisely, the first term refers to the case where the overall result on $s-1$ pairs as well as the result on the single 
pair are correct whereas for the second term Eve has failed in both cases.

Given that the probability $P_s(\textrm{suc}|{\bf k}, {\bf w})$ is a function of $s$ uncorrelated random 
variables $\omega_{j,n}$, its analysis for $s>2$ is rather cumbersome.  Nevertheless, it is straightforward to 
obtain an analytic expression for the average probability of success $\bar{P}_s(\textrm{suc})$ by averaging over 
all possible keys and codewords which is equivalent to averaging over all possible combinations of $\{\omega_{s,j}\}$. 
Along the lines of Appendix \ref{app2} it can be proven that
\bea
\bar{P}_s(\textrm{suc})=\frac{1}2+\frac{1}{2^{s+1}}.
\label{ind_ps}
\eea
Again, the average probability of success drops exponentially with increasing values of $s$. 
In contrast to Eqs. (\ref{Ubound}) and (\ref{swap}),  this expression does not depend on $T$ since the attack under 
consideration uses only one copy of the public key. It is, however, equivalent to the corresponding  expression for 
the forward-search attack, i.e. Eq. (\ref{swap}) for $T=1$. Hence, for a given security threshold $\varepsilon$ the 
length of the codeword has to satisfy inequality (\ref{Lbound3}) for $T=1$. 

\begin{table}[t]
\begin{center}
\begin{tabular}{|l|c|c|c|c|c|c|c|c|c|}
\hline
public key & t,t & t,f & t,t & t,f & f,t & f,f & f,t  & f,f\\
\hline
cipher state & t,t & t,f & f,f & f,t & t,f & t,t & f,t  & f,f\\
\hline
$e_1$,$e_2$ & 0,0 & 0,2 & 1,1 & 1,1 & 1,1 & 1,1 & 2,0  & 2,2\\
\hline
$e$ & 0 & 2 & 2 & 2 & 2 & 2 & 2  & 4\\
\hline
\end{tabular}
\caption{\label{tab:1} Encryption of a single bit, on the state of two qubits $(s=2)$. 
Possible combinations of true (t) and false (f) outcomes that lead to correct estimation of the 
message.}
\end{center}
\end{table}

\section{Conclusions}
We have analyzed the security of a quantum-public-key encryption (QPKE) scheme that relies on single-qubit rotations. 
For a given number of public keys  the symmetry underlying the protocol has been shown to restrict 
considerably the information  gain that an eavesdropper might gain on the private key. This result suggests that new 
more efficient QPKE schemes could rely on quantum one-way functions, which explore symmetries in the 
involved quantum states. It is also worth recalling here the pivotal role of symmetries in  
quantum-key-distribution protocols, as a result of which qudit-based protocols can tolerate higher error rates than
qubit-based ones \cite{qudit}.
  
The robustness of the protocol  under consideration was mainly analyzed in the framework of an attack which takes into account all the 
public-key copies and is based on projective measurements on single qubits. 
As a main result it has been shown that the performance of this attack is comparable to the performance of optimal collective 
measurements \cite{Der98} as well as to the forward-search attack of \cite{NikIoa09} which involves rather complicated 
quantum operations. Variants of the attack are expected to be applicable to other types of QPKE schemes as well.

\section*{Acknowledgements}
This work is supported by CASED. We are grateful to Joe Renes for useful suggestions 
and discussions. 

\begin{appendix} 
\section{Properties of the density operator (\ref{rho_tau}).}
\label{app1}
As for the matrix elements of the density operator of Eq. \eqref{rho_tau}, we can distinguish two different cases:
\\
{\em Case 1:} If $l+l^\prime$ is an even number, the function 
$f_{\tau,l}(k_j\theta_{n})f_{\tau,l^\prime}^\star(k_j \theta_{n})$ 
has even parity and does not change sign as we sum over all 
possible values of $k_j\in\Int_{2^n}$. 
Hence, we expect a non-zero contribution of $C_{l,l^\prime}$ in this case. 

{\em Case 2:} If $l+l^\prime$ is an odd number, the element $C_{l,l^\prime}$ vanishes 
since the parity of the overall trigonometric function in the sum is odd. 

Another important property of the density operator (\ref{rho_tau}) is that for fixed value of $\tau$
there seems to exist a critical value of $n$, let us say $n_{\rm c}$, for which 
it is $n$-independent for all $n\geq n_{\rm c}$. 
Furthermore, we have studied the rank of the density operator as well as the form of its eigenvalues
for various values of $n$ and $\tau$.  Our simulations show that for fixed $\tau$, $\rm{rank}[\rho_{j,\rm prior}^{(\tau)}]<\tau+1$ 
for all $n<n_{\rm c}$ and thus the density operator is singular, whereas for $n\geq n_{\rm c}$,  
$\rm{rank}[\rho_{j,\rm prior}^{(\tau)}]=\tau+1$.

The von Neumann entropy of a quantum state is bounded from above by $\log_2(D)$ with $D$ denoting the dimension of
the support of the relevant density operator. In view of the hermiticity of  $\rho_{j,\rm prior}^{(\tau)}$
we have $D=\textrm{rank}[\rho_{j,\rm prior}^{(\tau)}]$ and thus for a given pair of $(\tau,n)$
the entropy of the density operator is bounded from above by the corresponding entropy for $(\tau,n_{\rm c})$.
Hence, we arrive again at the upper bound for the entropy provided in (\ref{entropy_pri_1}).

In order to obtain a tighter bound we can investigate eigenvalues of the density operator for $(\tau,n_{\rm c})$.
Our simulations suggest that  in this case the eigenvalues of (\ref{rho_tau}) are given by 
\be
\lambda_i=\frac{1}{2^\tau}\binom{\tau}{i}. 
\ee
So, $S[\rho_{j,\rm prior}^{(\tau)}]$ can be calculated as 
the entropy of the binomial distribution with mean $\tau/2$ and variance $\tau/4$. This entropy is bounded from above by the entropy of the 
the normal (Gaussian) distribution with the same mean and variance \cite{book2}. Thus, we obtain the result
\be
\label{entropy_pri}
S[\rho_{j,\rm prior}^{(\tau)}]\leq\frac{1}{2}\log_2(\tau)+\frac{1}{2}\log_2(\pi e/2)
\ee
and this bound is below the one of (\ref{entropy_pri_1}). Accordingly, the information gain is upper bounded by
\be
I_{\rm av}\leq \frac{1}{2}\log_2(\tau)+\frac{1}{2}\log_2(\pi e/2).
\ee

\section{Proof of the upper bound (\ref{Ubound}).}
\label{app2}
The quantity we want to bound from above, i.e. $\bar{P}_s( \mathrm{suc})$, is a monotonously increasing function of  $\bar{P}( \mathrm{suc}|w_j)$ for $\bar{P}( \mathrm{suc}|w_j)>1/2$. 
Thus, in view of (\ref{U_of_T}) we have
\bea
\bar{P}_s( \mathrm{suc}) = \sum_{\substack{\alpha=0\\ {\rm even}}}^s \binom{s}{\alpha} [1-\bar{P}( \mathrm{suc}| w_j)]^{\alpha} 
[\bar{P}( \mathrm{suc} | w_j)]^{s-\alpha}\phantom{aa}\label{ap:eq1}\\
\leq \sum_{\substack{\alpha=0\\ {\rm even}}}^s \binom{s}{\alpha} [1-U(T)]^{\alpha} [U(T)]^{s-\alpha}.\phantom{aa}
\label{ap:eq2}
\eea
 
Let us denote the r.h.s of inequality (\ref{ap:eq2}) by $Q^{(s)}(T)$. 
It can be shown by induction that $Q^{(s)}$ is equal to (\ref{Ubound}). 
To this end we note that $Q^{(s)}$ can be written alternatively in the form of an iteration, i.e.
\bea
Q^{(s)}=Q^{(1)}Q^{(s-1)}+\left [1-Q^{(1)}  \right ] \left [1-Q^{(s-1)}\right ]. 
\label{Qi}
\eea 
For $s=1$ the equality we want to show holds, i.e. we have
\bea
Q^{(1)} = U(T)=\frac{1}2-\frac{1}{2}\left ( 1-\frac{1}{3T}\right ) := \frac{1}2+\frac{\lambda}2. 
\eea 
Assuming that  it holds for $s$, i.e.
\bea
Q^{(s)}=\frac{1}2+\frac{\lambda^s}2, 
\label{ap:eq3}
\eea 
we can prove also that it holds for $s+1$, because
\bea
Q^{(s+1)} 
 &=& \left(\frac{1}{2} + \frac{\lambda}{2}\right) \left(\frac{1}{2} + \frac{\lambda^s}{2}\right) + \nonumber\\
   & &+\left(\frac{1}{2} - \frac{\lambda}{2}\right) \left(\frac{1}{2} - \frac{\lambda^s}{2}\right)\\
&=& \frac{1}{2} + \frac{\lambda^{s+1}}{2}.
\eea

\end{appendix}


\begin{thebibliography}{99}
\bibitem{remark5}  For more information on the advantages of QPKE over protocols with point-to-point links the 
reader may look at chapter 1 of \cite{book1}.
\bibitem{remark7}
The $T^\prime$ copies of the public key that are available 
can be used for up to $T^\prime$ encryptions (one key per encryption). 
In general, multiple copies can be issued by the same user at the purpose of encryptions or 
eavesdropping.
\bibitem{remark6} 
The purpose of  the parity encoding is to randomize our encryption 
scheme \cite{NikIoa09}. The encoding is publicly known, and the security parameter 
$s$ has to be chosen judiciously so that security is guaranteed for a given number of 
public-key copies. Other types of randomization beyond parity encoding may be possible and equally efficient. 
A new security analysis of the protocol is required, if the accompanied encoding scheme differs 
from the one used here.   
Note also that analogous randomization techniques are used in conventional public-key encryption \cite{book1}.
\bibitem{remark1} In modern cryptography \cite{book1}, confidentiality (secrecy) and 
authenticity are considered as distinct and independent cryptographic goals
which are treated separately. Adopting the same attitude throughout this work 
we analyze the security of a particular public-key encryption scheme against certain 
types of attacks assuming that the possibility of an impersonation attack is prohibited 
by a reliable authentication scheme. The analysis of such an authentication scheme is beyond the scope of this 
work.
\bibitem{remark8}
Other encryption operations are also possible (e.g., mapping the different bit values on non-orthogonal qubit states), 
but they do not allow for a deterministic decoding. The following security analysis does not apply to this case.
\bibitem{remark2} Strategies that do not respect this symmetry are not expected to offer anything more.
\bibitem{remark3} This is essentially equivalent to the Schwinger representation 
of a spin-$\tau/2$ system in terms of two harmonic oscillators pertaining to 
$\ket{0_z}$ and $\ket{1_z}$.
\bibitem{remark9} The robustness of the protocol (with respect to security of the private key), may be increased considerably 
if $n$ is part of the private key and/or a random permutation is applied on the public qubits before they become publicly 
available \cite{Nik08}.
\bibitem{remark4} One public key has been issued by Bob for encryption of the message.
\end{thebibliography}
\end{document}